\documentclass[showkeys,preprintnumbers,amsmath,amssymb]{revtex4}
\usepackage{graphicx}
\usepackage{amssymb}
\usepackage{amsmath}
\usepackage{dcolumn}
\usepackage{bm}
\include{natbib}

\begin{document}
\title{A soft-lithographed chaotic electrokinetic micromixer for efficient chemical reactions in lab-on-chips}
\author{M.~Campisi}
\address{Department of Physics, University of North Texas, P.O.
Box 311427, Denton, TX 76203-1427, USA}
\author{D.~Accoto}
\thanks{Corresponding author}
\email{dino@sssup.it}\author{F.~Damiani}\author{P.~Dario}
\address{CRIM Lab Scuola Superiore Sant'Anna Viale R. Piaggio 34 56025 Pontedera
(Pisa) Italy}

\begin{abstract} Mixing is one of the basic functions which
automated lab-on-chips require for the effective management of
liquid samples. In this paper we report on the working principle,
design, fabrication and experimental characterization of a
soft-lithographed micromixer for microfluidic applications. The
device effectively mixes two liquids by means of \emph{chaotic
advection} obtained as an implementation of a Linked Twisted Map
(LTM). In this sense it is \emph{chaotic}. The liquids are
electrokinetically displaced by generating rolls through AC
electroosmosis on co-planar electrodes. The device performance has
been tested on dyed DI-water for several voltages, frequencies and
flow-rates, displaying good mixing properties in the range of $10
\div 100$kHz, at low peak-to-peak voltages ($\sim15 \div 20$
volts). Low voltage supply, small dimensions and possibility of
fabrication via standard lithographic techniques make the device
highly integrable in lab-on-a-chip platforms.
\end{abstract}


\maketitle

\section{\label{sec:intro}Introduction}
Since the beginning of the 1990s many research efforts have been
devoted to the development of integrated miniature fluidic systems
for the automated handling of biosamples on chip (lab-on-chip and
micro total analysis systems ($\mu TAS$)). In many biological and
chemical experimental procedures the ability to control the mixing
index between different reactants is often of paramount importance
for fast and efficient reactions. The development of micromixers
for microfluidic applications has attracted the attention of many
research groups worldwide, and a quite large number of papers
already appeared which report on the development of micromixers
based on different actuation strategies (see for instance
\cite{Campbell04}).

Mixing occurs when at least two species molecules spread around and diffuse uniformly in the bulk of a medium. Diffusion is a spontaneous process: the simplest micromixer is a simple channel into which two liquids are injected. If $D$ is the diffusion coefficient of a certain species, $w$ the characteristic dimension of the channel cross section, and $u$ the average flow speed, the Peclet number is $Pe=\frac{w}{D}u$ and the given
species mixes homogenously only after having travelled a distance over the channel $l\sim \frac{w^2}{D}u$. In order to reduce mixing times, channels with small cross sections are required. Nonetheless, the diffusive mixing length may be quite long. For example, mixing large macromolecular species, such as DNA or
proteins, which typically have a diffusion coefficient on the order
of $10^{-11} m^2 s^{-1}$ in a $200 \mu m$ wide channel, with an
average flow speed of $1 mm/s$, requires a channel length of about
$1 m$ with a corresponding mixing time exceeding $15 min$. This example shows that passive micromixers, which rely only on spontaneous diffusion of species suspended in liquid carriers, are not really compatible with miniature chips. On the contrary, active micromixers are able to both shortening mixing length and reducing mixing time by
``stirring'' the liquids, so that their interfacial area increases while the actual diffusion distance is greatly shortened.

Since flows in microchannels are typically controlled by viscosity
(``creeping flows'' at low Re numbers), stirring cannot be
achieved by means of turbulence. On the contrary, convective
transport can be effectively used to generate \emph{chaotic
advection}, that is streamlines stretching-and-folding in such a
way that originally close fluid particles follow trajectories
which diverge exponentially in time.

Also chaotic micromixers can be \emph{passive}
or \emph{active}. In the former class, \emph{chaotic
advection} may be induced by means of geometrical features of the
mixing channel (\cite{Liu00} and \cite{Stroock02}). In the latter class, active mixing is generated by means of an
external actuation which can be mechanical (e.g. in
\cite{Campbell04} self-assembled magnetic microstirrers are used),
or electrokinetic (e.g. \cite{Oddy01,Huang03,Wu05,Lee05}). The
second type is of greater practical interest because it does not involve
moving parts. On the other hand, current electrokinetic devices show a few drawbacks due to
the typically high electric voltages required. The device
presented here overcomes such limitations by taking advantage of
low-voltage AC electrokinetic rolls generation (see below).

The device was designed according to the Linked Twisted Map (LTM)
strategy described in \cite{Ottino04} and based on dynamic system
theory. As the liquids flow downstream, they encounter a number of
mixing elements which induce transverse circulating flows. As a
fluid particle transverse one element's length, its coordinates in
the cross-section plane change to new values according to a
\emph{map} of the cross section into itself. The flow through many
identical elements, then, can be modelled as the repeated
application of that map. The mixer would perform well only if the
map has effective mixing properties in the sense of dynamic system
theory. The LTM is one of the possible mixing maps which might be
implemented in real microfluidic mixers. The main idea of the LTM
is that of alternating an asymmetric (with respect to one of the
channel symmetry planes) circulating pattern with its symmetrical
one (with respect to the same plane) \cite{Ottino04}. The problem
has been studied mathematically \cite{Wiggins04}, to give the
conditions on the circulating flows' ``strength'' and ``overlap'',
which ensure the map to be mixing. The device proposed in
\cite{Stroock02}, is a striking example of chaotic advection
induced by an LTM. In that device, also named \emph{staggered
herringbone micromixer} the circulating flows are generated by
means of opportunely defined grooves on the top of the channel
and, as such, it is a passive device.

Recently\footnote{While this work was already under preparation.},
Sasaki et al. \cite{Sasaki06} have demonstrated that
electrokinetic rolls can be used for developing an active
micromixer. In this paper we further develop the approach
presented in \cite{Sasaki06} in the framework of the chaotic
LTM-mechanism and we present a novel device which differs from
that of \cite{Sasaki06} both in geometry and fabrication
technology. In particular, we show that the device can be
effectively fabricated using glass an polymers, instead of
silicon, using a low-cost rapid prototyping techniques called
soft-lithography \cite{Xia98}. Electrodes have been fabricated by
sandwiching Ti-Au-Ti layers, the first one (Ti) serving as an
adhesion layer over glass, the second one (Au) to reduce the
electrical impedance and the third one (Ti) as a dielectric layer,
since Ti spontaneously form a coherent dielectric oxide layer.
Compared to the device described in \cite{Sasaki06}, current mixer
performed well with very high Peclet numbers in the range
${0.6}*10^{6}$ - ${1.8}*10^{6}$.

In Sec. \ref{sec:Design} we report on the design of the
micromixer, whose fabrication is described in Sec.
\ref{sec:fabrication}. The last paragraphs are devoted to
described the experimental measurements carried on the device
(Sec. \ref{sec:Experimental}) and to discussion and conclusion
(Sec. \ref{sec:results}).

\section{\label{sec:Design}Design}
\begin{figure}[!]
  \includegraphics[width=12cm]{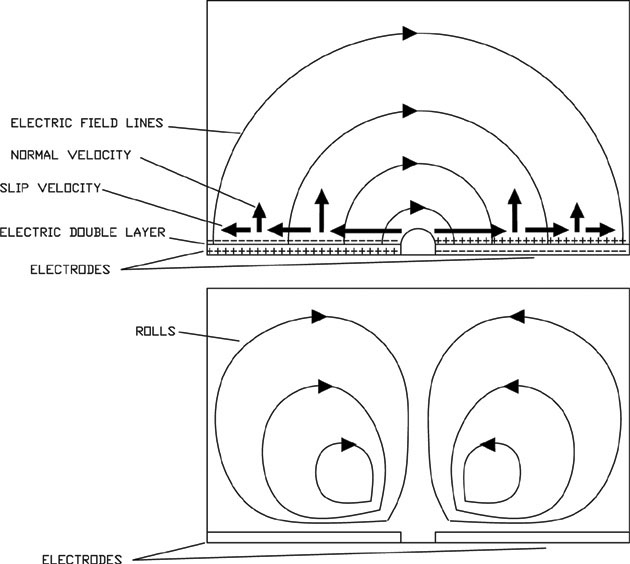}
  \caption{Schematic of electric field lines, slip and tangential velocities (top) and rolls (bottom)
   on top of planar electrodes addressed by an AC voltage.}
  \label{fig:rolls}
\end{figure}
\begin{figure}[!]
  \includegraphics[width=12cm]{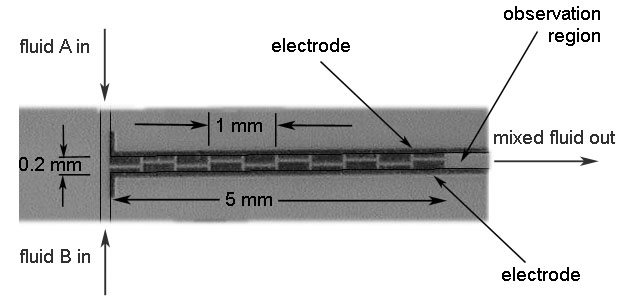}
  \caption{Microphotograph of the micromixer. Black solid lines have been superimposed
  for a better visualization of the T-channel.}
  \label{fig:mixing-element}
\end{figure}
The generation of rolls by means of AC electroosmosis has been
thoroughly studied both experimentally and theoretically
\cite{Green00,Gonzalez00,Green02,Ajdari95} and it is becoming
increasingly popular in the microfluidics community. Examples of
the employment of electrokinetic rolls include micropumps
\cite{Studer02} and control of microparticles \cite{Tuval05}.
 The mechanism of electrokinetic roll generation is
schematically illustrated in Fig. \ref{fig:rolls}. Two coplanar
micro-electrodes, spaced by a thin gap (on the order of $10 \mu m$
wide) are addressed with an AC low voltage ($\lesssim 20$ volts).
The electric field interacts with a ionic diffuse layer, which
forms on top of the electrodes, giving rise to a constant slip
velocity which points from the electrode gap outwards (see Fig.
\ref{fig:rolls}, top). Since the liquid is confined in a closed
cavity, the slip-velocity results in the formation of a couple of
counter-rotating rolling patterns (see Fig. \ref{fig:rolls},
bottom). We used asymmetric microelectrodes to produce a sequence
of asymmetric rolls in order to implement a LTM. The device
comprises a T-channel with two inlets for the fluids to be mixed
and one outlet for the mixed flow. The flow is pressure driven. As
the two liquids enter the main channel their flow is laminar
($Re<5$) with a very slow spontaneous mixing rate. The mixing
tract, hosting the $2$ micro-electrodes on its floor, starts at
entrance of the main channel. The electrodes are patterned so as
to define $5$ mixing elements (see Fig \ref{fig:mixing-element}),
each comprising two zones. In the first zone the gap ($20 \mu m$
wide) between the electrodes is placed at a distance $w/3$ from
one channel side and $2w/3$ from the other side, where $w=200 \mu
m$ is the channel width. In the second zone the gap is located
symmetrically with respect to the channel axis, so that the
positions of the two electrodes are interchanged (twisted).
Therefore the electrode gap has the shape of a square wave moving
along the channel axis. Each mixing element generates two couples
of asymmetrical rolls, one for each zone. The rolls generated in
the second zone are identical to those generated in the first
zone, except for their interchanged (twisted) location. Such
spatial twisting is at the basis of the implemented LTM strategy.

\begin{figure}
  \includegraphics[width=12cm]{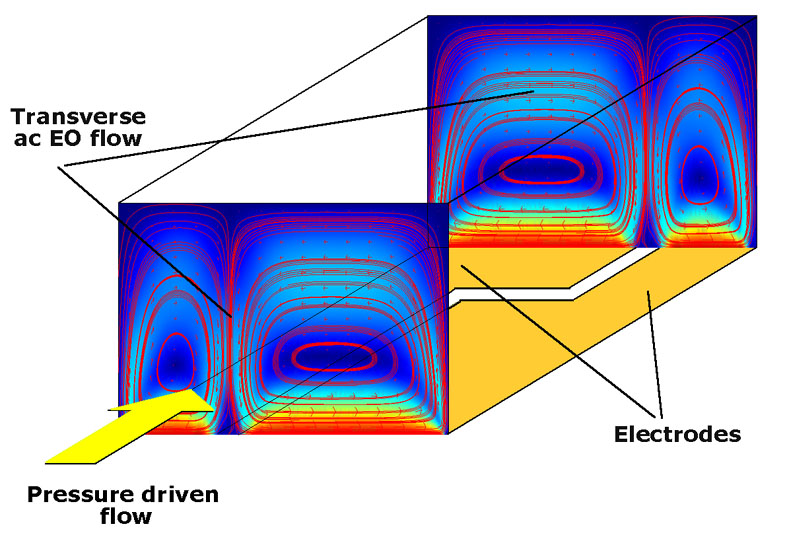}
  \caption{Schematic representation of circulating transverse flows (rolls) generated by AC electroosmosis which
  superimpose to the axial pressure driven flow within the micromixer. The positions of the asymmetric vortexes are
   twisted along the channel length in such a way as to implement a linked-twisted-map.}
  \label{fig:LTM}
\end{figure}

Because of the very low Re number, the resulting flow in the
channel can be imagined as the superimposition of an axial
pressure driven flow with transversal and twisted circulating
flows, as schematically represented in Fig. \ref{fig:LTM}.
\begin{figure}
  \includegraphics[width=8cm]{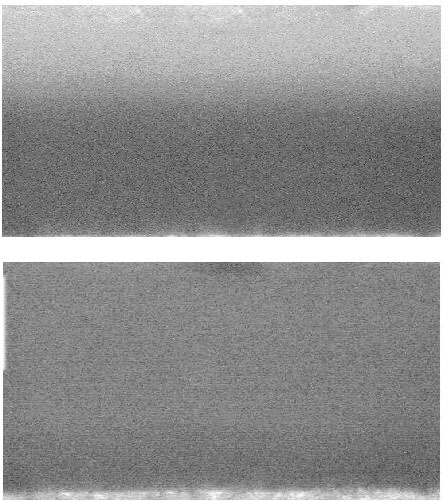}
  \caption{Normalized pictures of unmixed streams (top) and mixed streams (bottom) in the micro-channel.
  The images show the region of the channel right downstream with respect to the mixing region.
  The device was turned off when the top image was shot and was turned on at $100 kHz$, $20 V_{p-p}$ in the bottom image.
  The Peclet number was ${0.58}\times 10^{6}$.}
  \label{fig:mix-pictures}
\end{figure}
\begin{figure}
  \includegraphics[width=12cm]{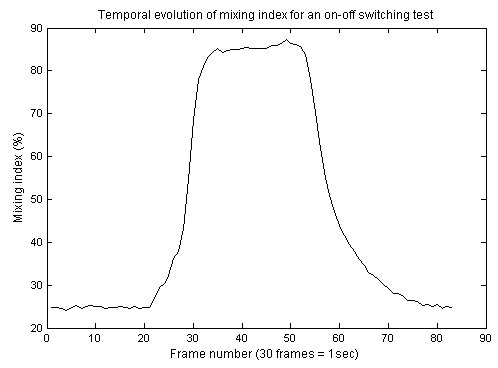}
  \caption{Typical temporal evolution of mixing index as the device is turned on and off.}
  \label{fig:time-mix1}
\end{figure}
\begin{figure}[!]
  \includegraphics[width=10cm]{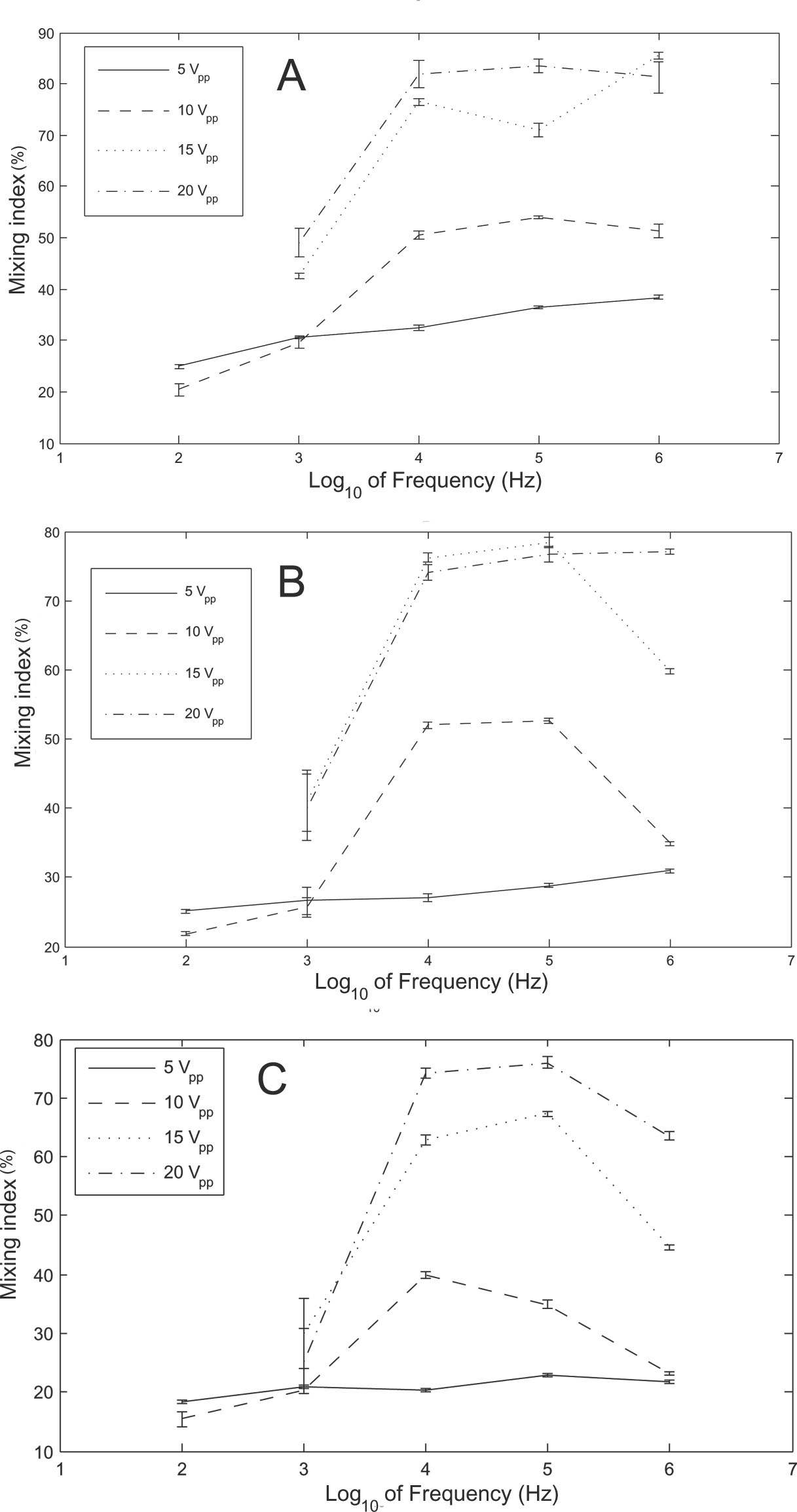}
  \caption{Mixing performance as a function of driving frequency for different peak-to-peak voltage amplitudes at
  $Pe ={0.58}*10^{6}$ (A), $Pe ={1.22}*10^{6}$ (B), $Pe ={1.86}*10^{6}$ (C). Data corresponding to a driving voltage of $100 Hz$ at $V_{pp}\geqq 15V$ are not reported, being affected by bubbles generated by water electrolysis.}
  \label{fig:mix-volt2}
\end{figure}
The device might also work with other periodic shape of
electrodes. As an example, in \cite{Sasaki06} a device with a
sinusoidal gap between electrode is presented.

\section{\label{sec:fabrication}Fabrication}
The channel has been fabricated in PDMS using soft-lithographic
techniques \cite{Xia98}. The electrodes where patterned on a glass
substrate using a lift-off process. The electrodes were made of an
Au layer ($\sim 100 nm$) sandwiched between two Ti layers ($\sim
10 nm$ bottom, $\sim 20 nm$ top) as described in \cite{Green00}.
The bottom layer serves as an adhesion layer, whereas the top
layer serves for insulation purposes (Titanium develops an
insulating superficial thin oxide layer at room temperature and
pressure conditions). The PDMS channel was aligned and mounted on
the patterned glass substrate to define the micro-device. Sealing
between PDMS and glass was obtained by assembly the parts in a
purposely developed mechanical support. The main channel is $200
\mu m$ wide and $50 \mu m$ high. The mixing region is $5 mm$ long
for a total of five $1 mm$ long mixing elements.

\section{\label{sec:Experimental}Experimental}
The micro-device was treated in an oxygen plasma ($3 min$, $18 W$)
to make the channels hydrophilic, then it was primed with DI
water. The two inlet reservoirs were respectively loaded with DI
water and methylene blue dyed water. The pressure was applied by
connecting the inlets to a pressurised air chamber. The pressure
was measured using a digital manometer (model 840081 by Sper
Scientific). The pressure in the air chamber was generated
hydrostatically connecting the chamber to a vertical column filled
with water. The device outlet was left at room pressure ($1 atm$).

The electrodes were connected to a wave form
generator with $20 V_{p-p}$ maximum output voltage and addressed
with a sinusoidal wave. In the experimental tests we observed the region of the channel
which follows the mixing region using a CCD microscope (from Hirox
inc.). Digital videos at 30 frames per second were stored in a PC equipped with a frame-grabber
(VCE-pro by Imperx Inc.) for off-line analysis.

Figure \ref{fig:mix-pictures} shows two pictures ($300 \times 550$
pixels) taken at constant applied relative pressure ($0.25 psi$,
corresponding to $Pe={0.58} \times 10^{6}$) when the device is off
(top) and on. In both cases the applied voltage is $20 V_{pp}$ at
$100 kHz$. The picture shows how the blue dye spreads
homogeneously through the channel cross section.

To quantify the mixing level starting from images analysis we
adopted the standard deviation method as described in
\cite{Stroock02}. The frames were first converted to grey-scale
and then pixel values $I$ were normalized to $1$, so that $0$
corresponds to white and $0$ to black. Finally, standard deviation
$\sigma = <I^2-<I>^2>^{1/2}$ of pixel values was computed.

Two theoretical limit cases exist. For perfectly unmixed
fluids\footnote{In practice, diffusion always creates a finite
grey region at interface.} half pixels are black and half are
white. In this case it is $\sigma= 0.5$. For fully mixed fluids
all the pixels have the same grey color and it is $\sigma = 0$.
Based on this observation, we define the mixing index as:
\begin{equation}\label{mix-index}
    m = 1-2\sigma,
\end{equation}
in such a way that $m=0$ corresponds to completely unmixed fluids and
$m=1$ to completely mixed ones. It is worth underlining that this
definition of mixing index contains an element of arbitrariness as
it depends on the contrast of the image. Further, since the
standard deviation is used, $m$ is affected by the CCD
noise. As a consequence, images
need to be pre-processed before computation of the standard
deviation. We proceeded as follow. Since the mixing index $m$ is evaluated over the cross sections of the channel, which correspond to lines perpendicular to the axis as viewed in the grabbed frames, pixels intensities have been averaged along rows, so as to reduce each $300 \times 550$ image
matrix to a $300 \times 1$ array. This array contains the mean value of
pixel intensities along each row. To overcome the arbitrariness due to contrast
we took the first frame acquired during the experimental
session, reduced it to an array as explained above, evaluated its
maximum and minimum pixel values $I_{min}$, $I_{max}$ and adjusted
the contrast by linearly mapping the pixel value interval
$[I_{min},I_{max}]$ into the interval $[0,1]$. The same
transformation was then applied to all the $300 \times 1$ arrays
obtained from the frames acquired during the experimental session.
Finally the mixing index was calculated for each array.

We performed a number of experimental tests at different applied
pressure differences ($\Delta P =  0.25, 0.50, 0.75$ psi,
corresponding respectively to measured Peclet numbers of $Pe =
{0.58}*10^{6}, {1.22}*10^{6}, {1.86}*10^{6}$), for different
values of driving frequency (ranging from $10^2$ to $10^6$ Hz) and
four values of peak-to-peak voltage amplitude ($5$,$10$, $15$ and
$20$ volts). The chosen Peclet numbers correspond to average flow
velocities of $29$, $61$ and $93 \; mm/sec$ and flow rates of
$0.29, 0.61, 0.93 \mu l/s$.

The tests consisted in turning the device on and off while
recording a video. Fig. \ref{fig:time-mix1} shows a typical
temporal evolution of the mixing index. For each test we computed
the temporal evolution of the mixing index. The average mixing
index calculated over the high plateau has been taken as a measure
of the mixing performance of the device at the corresponding
values of frequency, voltage and flow-rate. The error on this
measure is given by the standard deviation of the mixing index on
the plateau.

\section{\label{sec:results}Results and discussion}
The results of all tests are collected in Fig.
\ref{fig:mix-volt2}. The mixing index is higher for higher applied
voltages. All the curves display a similar behavior: the mixing
index grows with frequency up to the value of $10$ kHz, where a
plateau begins. In most cases a decrease is observed around
$1Mhz$, where electrothermal phenomena may occur. Peak performance
($m \simeq 0.85$) is observed at frequencies between $10$ kHz and
$1$ MHz, for the lowest Peclet number employed (${0.58}*10^{6}$).
The presence of a peak frequency around 100 kHz was expected on
the basis of the theoretical analysis of Ref.
\cite{PhysRevE.61.R45}: at lower frequencies the field quickly
relaxes into the bulk, whereas at higher frequencies the electric
charges do not have time to build up over electrodes.

Measurements also show that the mixing index decreases with increasing flow rates, so that longer mixing lengths are required for higher flow velocities. This behavior is expectedly typical for such devices and a similar one was also reported in \cite{Sasaki06}.

We conclude by observing a few advantages and drawbacks of this
device with respect to other similar ones reported in literature.
For what concerns the drawbacks we should notice that, unlike the
\emph{staggered herringbone micromixer} \cite{Stroock02}, the
axial and transverse flows are independent from each-other.
Therefore, as the axial velocity increases, one has to increase
the velocity of the circulating flows accordingly in order to
guarantee that the resulting map be mixing. Practically this poses a
limit to the range of flow rates over which the device performs well.
Nonetheless, the devised application as micromixer for Lab-on-Chip is not hindered by such limit, since the limit flow-rates are above those which can practically be produced in
Lab-on-a-chip devices.

Among the advantages we should mention that unlike other
electrokinetic micro-mixers \cite{Oddy01,Wu05,Lee05} which use
high voltages ($100 \div 1000$ volts), here the voltage employed
is quite low ($\lesssim 20$ volts). Furthermore, the high
frequency AC operation preserve the device from observable
electrolytic bubbles generation.

Another nice feature of
this device is its scalability. As the channel cross section gets
smaller it is expected that the same performances be reached with
lower voltages. The main advantage of active micromixing is not
only that the liquid can be stirred at will, but it can be mixed
up to a desired level by tuning the applied voltage. This makes
the device amenable to be integrated with a programmable control
system which records pixel values with a CCD sensor, process them
in real time and tunes the voltage output in order to control the
mixing level.

All these properties, along with the small dimensions and the
possibility, which we have demonstrated, of using soft-lithography as fabrication technique, makes the device ideal for
integration on low-cost, multi-functional Lab-on-a-chip platforms.
\newpage
\section*{Acknowledgements}
This work has been supported by the
Italian Ministry of Education, University and Research through
grants FIRB RBNE01T22-003 and FIRB RBNE01ZB7A-005.

\end{document}